\documentclass[runningheads]{llncs}
\usepackage[T1]{fontenc}
\usepackage[utf8x]{inputenx}

\usepackage[noadjust,nospace]{cite}
\usepackage{xspace}
\usepackage[hyphens]{url}
\usepackage[breaklinks=true]{hyperref}
\usepackage{upgreek}
\usepackage{multirow}
\usepackage{amssymb}
\usepackage{amsmath}
\usepackage{cleveref}
\usepackage{booktabs}
\usepackage{wrapfig}
\usepackage{color}
\usepackage{tikz}
\usepackage{ dsfont }
\usepackage{ifthen}

\Crefname{figure}{Fig.}{Figs.}
\crefname{figure}{fig.}{figs.}
\Crefname{tabular}{Tab.}{Tabs.}
\crefname{tabular}{tab.}{tabs.}
\Crefname{section}{Sect.}{Sects.}
\crefname{section}{sect.}{sects.}
\Crefname{equation}{Eq.}{Eqs.}
\crefname{equation}{eq.}{eqs.}
\Crefname{definition}{Def.}{Defs.}
\crefname{definition}{def.}{defs.}
\Crefname{theorem}{Thm.}{Thms.}
\crefname{theorem}{thm.}{thms.}

\newboolean{hideplots}
\newlength{\backopssize}
\usepackage{pgfplots}
\pgfplotsset{compat=1.14}
\usetikzlibrary{calc,angles,quotes,automata,arrows,decorations.pathmorphing,patterns,positioning,fit,trees,shapes,shapes.misc,shadows}

\newcommand{\colorpar}[3]{\colorbox{#1}{\parbox{#2}{#3}}}
\newcommand{\marginremark}[3]{\marginpar{\colorpar{#2}{\linewidth}{\color{#1}#3}}}
\newcommand{\todo}[1]{\textcolor{red}{\texttt{TODO:} #1}}
\newcommand{\jpk}[1]{\marginremark{white}{green!70!black}{\scriptsize{[JPK]~ #1}}}
\newcommand{\js}[1]{\marginremark{white}{orange}{\scriptsize{[JS]~ #1}}}
\newcommand{\bk}[1]{\marginremark{white}{blue!70!black}{\scriptsize{[BK]~ #1}}}
\newcommand{\ah}[1]{\marginremark{white}{red}{\scriptsize{[AH]~ #1}}}

\newcommand{\tool}[1]{\textsc{#1}\xspace}
\newcommand{\toolset}{\tool{Modest Toolset}}

\renewcommand{\emptyset}{\ensuremath{\varnothing}}

\newcommand{\ie}{i.e.\ }

\newcommand{\wrt}{w.r.t.\xspace}

\newcommand{\set}[1]{\ensuremath{\{\,#1\,\}}}
\newcommand{\tuple}[1]{\ensuremath{\langle #1 \rangle}}

\newcommand{\defeq}{\mathrel{\vbox{\offinterlineskip\ialign{\hfil##\hfil\cr{\tiny \rm def}\cr\noalign{\kern0.30ex}$=$\cr}}}}
\newcommand{\True}[0]{\ensuremath{\mathit{true}}\xspace}
\newcommand{\False}[0]{\ensuremath{\mathit{false}}\xspace}

\newcommand{\until}{\ensuremath{\mathbin{\mathsf{U}}}}

\newcommand{\enab}[1]{\ensuremath{\mathit{enab}\ifx#1\empty#1\else{(\ensuremath{\mathit{#1}})}\fi}\xspace} 

\newcommand{\PTA}[1][]{\ensuremath{\mathcal{P}_{#1}}\xspace} 

\newcommand{\solprop}[3]{\ensuremath{\mathds{P}_{#1#2}(\lozenge{\,#3})}}

\newcommand{\nodemdp}[1][node distance=4cm]{\node[draw, circle, fill=black, minimum size=1mm, scale=0.5, {#1}]}

\Crefname{figure}{Fig.}{Figs.}
\crefname{figure}{fig.}{figs.}
\Crefname{tabular}{Tab.}{Tabs.}
\crefname{tabular}{tab.}{tabs.}
\Crefname{section}{Sect.}{Sects.}
\crefname{section}{sect.}{sects.}
\Crefname{equation}{Eq.}{Eqs.}
\crefname{equation}{eq.}{eqs.}
\creflabelformat{equation}{#2#1#3}

\makeatletter%
\g@addto@macro\normalsize{%
  \setlength\abovedisplayskip{3pt}%
  \setlength\belowdisplayskip{3pt}%
  \setlength\abovedisplayshortskip{-3pt}%
  \setlength\belowdisplayshortskip{3pt}%
}%
\makeatother

\makeatletter
\def\orcidID#1{\smash{\href{http://orcid.org/#1}{\protect\raisebox{-1.25pt}{\protect\includegraphics{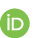}}}}}
\makeatother

\hypersetup{
  pdftitle = {Backwards Reachability for Probabilistic Timed Automata: A Replication Report}
}

\begin{document}

\title{%
Backwards Reachability for Probabilistic\\Timed Automata: A Replication Report%
\thanks{%
Authors are listed in alphabetical order.
This work was funded by
NWO grant OCENW.KLEIN.311,
NWO VENI grant 639.021.754,
and the EU's Horizon 2020 research and innovation programme under MSCA grant agreement no.\ 101008233.
}%
}
\titlerunning{Replicating Backwards Reachability for Probabilistic Timed Automata}

\author{
Arnd Hartmanns\,\orcidID{0000-0003-3268-8674}
\and Bram Kohlen\,\orcidID{0000-0003-2908-8838}
}
\authorrunning{Hartmanns, Kohlen}

\institute{
University of Twente, Enschede, The Netherlands
}
\maketitle

\begin{abstract}
Backwards reachability is an efficient zone-based approach for model checking probabilistic timed automata \wrt PTCTL properties.
Current implementations, however, are restricted to maximum probabilities of reachability properties.
In this paper, we report on our new implementation of backwards reachability as part of the \toolset.
Its support for minimum and maximum probabilities of until formulas makes it the most general implementation available today.
We compare its behaviour to the experimental results reported in the original papers presenting the backwards reachability technique.
\end{abstract}

\section{Introduction}
\label{sec:Introduction}
Probabilistic timed automata (PTA)~\cite{KNSS02} combine the capabilities of Markov decision processes (MDPs)~\cite{P94} to model decision-making under uncertainty with the real-time features of timed automata (TA)~\cite{AD94}.
As the state space of a PTA is uncountably infinite, model checkers transform PTA to equivalent finite-state models.
The region graph~\cite{KNSS02}, zone-based forwards reachability~\cite{KNSS02}, zone-based backwards reachability~\cite{KNSW07}, and digital clocks~\cite{KNPS06} transformations map to MDPs.
Since region graphs are too large, and the forwards approach merely delivers upper bounds on maximum reachability probabilities, only the latter two approaches are relevant today.
Although limited to closed clock constraints and reachability properties, the digital clocks approach is easy to implement and also supports expected-reward properties.
Backwards reachability supports the full logic PTCTL; and while its complexity is doubly exponential in the number of clocks, it tends to be efficient in practice and competitive with digital clocks~\cite{KNP09}.
The game-based abstraction technique~\cite{KNP09} maps to stochastic games.

In this paper, we focus on the backwards reachability approach:
it poses no restrictions on the PTA, supports all of PTCTL, and maps to MDPs.
The latter is favourable for methods like parametric model checking that currently work on MDPs only~\cite{HKKS21}.
However, there is currently no tool that fully implements backwards reachability.
PRISM's~\cite{KNP11} backwards engine is restricted to maximum probabilities and the ``eventually'' operator (lacking support for ``until''), and does not support models with global variables.
The prototype implementation of~\cite{KNSW07} also supported minimum probabilities, but is no longer available.

In this paper, we report on our replication of backwards reachability in the \toolset~\cite{HH14} that computes minimum and maximum probabilities of until formulas.
We compare its behaviour to the original prototype based on the experiments of~\cite{KNSW07} and its earlier conference version~\cite{KNSW04}.
While we obtain the same probabilities, our implementation's behaviour and performance is notably different.
We in particular found that the original papers omit the algorithm for timed predecessors, which is central to the approach, and rather involved.

\section{Probabilistic Timed Automata}
\label{sec:Preliminaries}
\begin{wrapfigure}[10]{r}{4cm}
	\vspace{-5ex}
	\resizebox{\linewidth}{!}{
	\begin{tikzpicture}[every node/.style={rounded corners=5pt, node distance=4cm, font=\large},simpstate/.style={draw,fill=white,inner sep=1pt,minimum width=14pt},baseline=(s0)]
		\node[draw, initial, initial text={}] (s0) {\shortstack{$\mathstrut{}\mathit{init}$ \\ $x \leq 2 \,\land$\\ $y \leq 24$}};
		\node[draw] (s1) [right of=s0] {\shortstack{$\mathstrut{}\mathit{lost}$ \\ $x \leq 8$}};
		\node[draw] (s2) [below of=s0] {\shortstack{$\mathstrut{}\mathit{fail}$ \\ $\True$}};
		\node[draw] (s3) [below of=s1] {\shortstack{$\mathstrut{}\mathit{done}$ \\ $\True$}};
		
		\nodemdp (s0a) [below right of=s0, yshift=-2.5cm, xshift=2.5cm] {};
		\nodemdp (s0b) [below of=s0, yshift=-0.5cm] {};
		\nodemdp (s1a) [left of=s1, xshift=0.15cm] {};
		\draw[-] (s0) edge node[left, align=right] {$\mathit{t\_out}$,\\ $ y \geq 18$} (s0b);
		\draw[->] (s0b) edge node[left] {$\emptyset, 1$} (s2);
		\draw[-] (s0) edge node[above, sloped] {$\mathit{send}, x \geq 1$} (s0a);
		\draw[->] (s0a) edge node[right] {$\emptyset, 0.1$} (s1);
		\draw[-] (s1) edge node[above,overlay] {\shortstack{$\mathit{retry}$,\\$x = 8$}} (s1a);
		\draw[->] (s1a) edge node[above] {\shortstack{$\{x\}, 1$}} (s0);
		\draw[->] (s0a) edge node[above, xshift=0.3cm] {$~\emptyset, 0.9$} (s3);
		\draw[->] (s2) edge[loop, in=-60, out=-120, min distance=0.75cm]
			node[pos=0.5, below, xshift=-2em] {$f, \True$}
			node[pos=0.5, draw, circle, fill=black, minimum size=1mm, scale=0.5] {}
			node[pos=0.5, below, xshift=1.5em] {$\emptyset, 1$}
			(s2);
		\draw[->] (s3) edge[loop, in=-60, out=-120, min distance=0.75cm]
			node[pos=0.5, below, xshift=-2.25em] {$d, \True$}
			node[pos=0.45, draw, circle, fill=black, minimum size=1mm, scale=0.5] {}
			node[pos=0.5, below, xshift=1.25em, overlay] {$\emptyset, 1$}
			(s3);
	\end{tikzpicture}
	}
\end{wrapfigure}

We show an example PTA $\PTA$ representing a basic communication protocol with message loss on the right. 
Its initial \emph{location} is $\mathit{init}$; it has two clocks $x$ and $y$.
The state of a PTA pairs the current location $\ell$ with the current \emph{clock valuation} $v$, which maps every clock to its non-negative real value.
All paths start in $\tuple{\mathit{init}, v_0}$ where $v_0$ assigns value $0$ to $x$ and $y$.
Letting $t > 0$ time units elapse in a location $\ell$ corresponds to the transition from $\tuple{\ell,v}$ to $\tuple{\ell,v+t}$ where $(v+t)(z) = v(z)+t$ for each clock $z$.
Every location has an \emph{invariant} constraining the passage of time:
time can only pass while the invariant remains satisfied.
The invariant of $\mathit{init}$ is $x \leq 2 \land y \leq 24$.
Location $\mathit{init}$ has two \emph{edges} labelled with actions $\mathit{send}$ and $\mathit{t\_out}$.
Picking action $\mathit{send}$ takes us to $\mathit{done}$ with probability $0.9$ and to $\mathit{lost}$ with probability $0.1$. 
Action $\mathit{send}$ is only available when the edge's \emph{guard}, $x \geq 1$, is satisfied.
Action $retry$ brings us back to $\mathit{init}$ with probability $1$ and \emph{resets} the value of $x$ to zero.

Due to the delay transitions, a PTA's state space is uncountably infinite.
To make the analysis of PTA tractable, we group clock valuations satisfying a constraint into \emph{zones}.
For example, the zone over constraint $x \leq 2 \land y \leq 24$ contains $v$ with $v(x) = 1.5$ and $v(y) = 23.9$, but not $v'$ with $v'(x) = 1.5$ and $v'(y) = 24.1$.
Given zone $\zeta$, the \emph{symbolic state} $\tuple{\ell,\zeta}$ contains all states $\tuple{\ell,v}$ where $v \in \zeta$. 
From now on, we use the notation for constraints to denote zones, too. 
A common data structure to represent a zone is the \emph{difference bound matrix} (DBM)~\cite{D89}. 
A DBM's entries are pairs of an integer and an operator $\lesssim\ \in \set{<, \leq}$. 
Entry $A_{i,j} = \tuple{d, \lesssim}$ means that $x_i - x_j \lesssim d$, given a bijection between the PTA's $n$ clocks and $x_1, \ldots, x_n$.
$x_0$ represents the clock that always has value zero.

We focus on model checking PTA \wrt \emph{until formulas}.
Property $\neg \mathit{avoid} \until \mathit{target}$ characterises the paths that eventually reach a state satisfying formula $\mathit{target}$ over locations and clock constraints without entering an $\mathit{avoid}$ state before. 
If $\mathit{avoid}$ is $\False$, then we have a \emph{reachability} formula of the form $\lozenge{\mathit{target}}$.
The properties we aim to check are statements about the probability of the set of paths characterised by a formula being smaller or larger than a given bound.
For example, property $\solprop{\geq}{0.99}{\mathit{done}}$ tests whether the probability to eventually reach location $\mathit{done}$ for \emph{any} choice of actions is greater than or equal to $0.99$. 
That is, the \emph{minimal probability} $\mathbb{P}_\mathrm{min}$ must be at least $0.99$. 
This is true for $\PTA$, as it must perform $send$ at least twice before $t\_out$ is enabled.
Conversely, $\solprop{\leq}{0.99}{\mathit{done}}$ states that the \emph{maximal probability} $\mathbb{P}_\mathrm{max}$ to reach location $\mathit{done}$ is at most $0.99$. 
This is false for $\PTA$ it is possible to perform $send$ three times which has a probability of $0.999$ to reach $done$.
However, $z.\solprop{\leq}{0.99}{\mathit{done} \land z \leq 10}$, where $z$ is a \emph{property clock}, is true as we can only send twice in 10 time~units.

\section{Backwards Reachability}
\label{sec:Backwards}
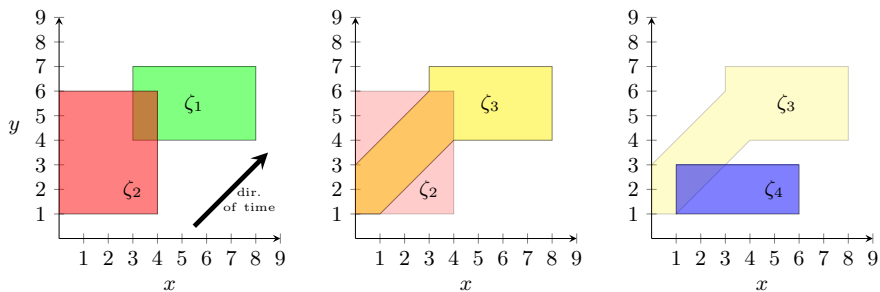
\begin{figure}[t!]
\centering
\scalebox{0.9}{
\begin{tikzpicture}
	\begin{axis}[
		width=\backopssize,
		height=\backopssize,
	    axis lines=middle,
	    	every axis x label/.style=
				{at={(ticklabel cs: 0.5,0)}, anchor=north},
			every axis y label/.style=
				{at={(ticklabel cs: 0.5,0)}, anchor=east},
			xmin=0,xmax=9,ymin=0,ymax=9,
			xtick distance=1,
			ytick distance=1,
			xlabel=$x$,
   			ylabel=$y$,
			title={}
	]
		\draw [fill=green, opacity=0.5] (3,4) rectangle (8,7);
		\draw [fill=red, opacity=0.5] (0,1) rectangle (4,6);
		\node [text width=0.3cm] at (5.5, 5.5) {$\zeta_1$};
		\node [text width=0.3cm] at (3, 2) {$\zeta_2$};
		
		\draw [-stealth,line width=2pt](5.5,0.5) -- (8.5,3.5);
		\node [text width=0.2cm] at (7.5, 1.9) {\tiny dir.};
		\node [text width=0.6cm] at (7.5, 1.3) {\tiny \text{of time}};
	\end{axis}
\end{tikzpicture}
}~
\scalebox{0.9}{
\begin{tikzpicture}
	\begin{axis}[
		width=\backopssize,
		height=\backopssize,
	    axis lines=middle,
	    	every axis x label/.style=
				{at={(ticklabel cs: 0.5,0)}, anchor=north},
			every axis y label/.style=
				{at={(ticklabel cs: 0.5,0)}, anchor=east},
			xmin=0,xmax=9,ymin=0,ymax=9,
			xtick distance=1,
			ytick distance=1,
			xlabel=$x$,
			title={}
	]
		\draw [fill=yellow, opacity=0.5] (8,7) -- (8,4) -- (4,4) -- (1,1) -- (0,1) -- (0,3) -- (3,6) -- (3,7) -- cycle;		
		
		\draw [fill=red, opacity=0.2] (0,1) rectangle (4,6);
		\node [text width=0.3cm] at (5.5, 5.5) {$\zeta_3$};
		\node [text width=0.3cm] at (3, 2) {$\zeta_2$};
	\end{axis}
\end{tikzpicture}
}~
\scalebox{0.9}{
\begin{tikzpicture}
	\begin{axis}[
		width=\backopssize,
		height=\backopssize,
	    axis lines=middle,
	    	every axis x label/.style=
				{at={(ticklabel cs: 0.5,0)}, anchor=north},
			every axis y label/.style=
				{at={(ticklabel cs: 0.5,0)}, anchor=east},
			xmin=0,xmax=9,ymin=0,ymax=9,
			xtick distance=1,
			ytick distance=1,
			xlabel=$x$,
			title={}
	]
		\draw [fill=yellow, opacity=0.2] (8,7) -- (8,4) -- (4,4) -- (1,1) -- (0,1) -- (0,3) -- (3,6) -- (3,7) -- cycle;		
		
		\draw [fill=blue, opacity=0.5] (1,1) rectangle (6,3);
		\node [text width=0.3cm] at (5.5, 5.5) {$\zeta_3$};
		\node [text width=0.3cm] at (5, 2) {$\zeta_4$};
	\end{axis}
\end{tikzpicture}
}
\caption{Time and discrete predecessors of zones.}
\label{fig:backops}
\end{figure}

The \emph{backwards reachability} approach transforms a PTA into an MDP via a backwards search starting at the $\mathit{target}$ symbolic state.
It then discovers those symbolic states that can reach $\mathit{target}$ by performing a transition and letting time pass. 
This procedure is iterated until all backwards-reachable states are discovered. 
We refer to this as the $\mathit{MaxU}$ algorithm~\cite[Fig.~5]{KNSW07}; it builds an MDP whose states are symbolic states of the PTA.
It needs to compute \emph{time predecessors} and \emph{discrete predecessors}.
The former operation, denoted $\mathit{tpre}_{\tuple{\ell,\zeta'}}(\tuple{\ell,\zeta})$, calculates the set of clock valuations that eventually reach $\zeta$ without leaving $\zeta'$.
In \Cref{fig:backops}, we visualise zones $\zeta_1 = (3 \leq x \leq 8 \land 4 \leq y \leq 7)$ and $\zeta_2 = (x \leq 4 \land 1 \leq y \leq 6)$ on the left.
In the middle, we show $\zeta_3 = \mathit{tpre}_{\tuple{\ell,\zeta_2}}(\tuple{\ell,\zeta_1})$.
The discrete predecessor operation, denoted $\mathit{dpre}(\tuple{\ell',a,X,\ell},\tuple{\ell,\zeta_3})$, makes a backwards transition over the specified edge and branch from $\ell'$ to $\ell$.
It performs a \emph{backwards reset} on the clocks in $X$.
On the right of \Cref{fig:backops}, we show $\tuple{\ell', \zeta_4} = \mathit{dpre}(\tuple{\ell', \alpha, \{x\}, \ell}, \tuple{\ell, \zeta_3})$ for invariant $x \leq 6$ of  $\ell'$ and guard $x \geq 1$ of $\alpha$.

$\mathit{MaxU}$ works for maximal probabilities only.
We restructure properties requiring minimal probabilities to properties with maximal probabilities:
To obtain, say, $\mathbb{P}_\mathrm{max}(\lozenge\,\mathit{target})$, we compute $1 - \mathbb{P}_\mathrm{max}(\neg\mathit{target} \until \phi)$ using $\mathit{MaxU}$ where $\phi = \mathbb{P}_\mathrm{max}(\square\neg\mathit{target}) \geq 1$.
The symbolic states satisfying $\phi$ can be precomputed via a graph analysis using the $\mathit{MaxV_{\geq 1}}$ algorithm~\cite[Fig.~7]{KNSW07}.
It iteratively removes those symbolic states that do not have a path inside $\neg\mathit{target}$ of duration $\geq c$, until a fixpoint is reached.
$\mathit{MaxV_{\geq 1}}$ is parameterised by $c$, which can be any positive integer; the value of $c$ influences the number of iterations and therefore the runtime.
The optimal value of $c$ is model- and property-dependent.
Each iteration of $\mathit{MaxV_{\geq 1}}$ in turn calls the $\mathit{MaxU_{\geq 1}}$ algorithm~\cite[Fig.~4]{KNSW07}, which calculates whether an until formula is satisfied with maximal probability~$1$.

\begin{figure}[t]
	\centering
	\resizebox{\linewidth}{!}{
		\begin{tikzpicture}[on grid,scale=1, node distance=4.5cm, nodestyle/.style={draw,circle,minimum size=1.9cm},baseline=(s0)] 
    
	   		\node [nodestyle] (s0) at (0,0) [on grid] {\shortstack{$s_4: lost$ \\ $x = 8$ \\ $y \leq 1$}};
			\node [nodestyle, initial above, initial text={}] (s1) [on grid, right of= s0] {\shortstack{$s_3: init$ \\ $1 \leq x \leq 2$ \\ $y - x \leq 1$}};
			\node [nodestyle] (s2) [on grid, right of= s1] {\shortstack{$s_2: lost$ \\ $x = 8$ \\ $y \leq 9$}};
			\node [nodestyle, initial above, initial text={}] (s3) [on grid, right of= s2] {\shortstack{$s_1: init$ \\ $1 \leq x \leq 2$ \\ $y \leq 10$}};
			\node [nodestyle] (s4) [on grid, right of= s3] {\shortstack{$s_0: done$ \\ $y \leq 10$}};

			\draw (s0) edge[->] node[pos=0.25, above] {\small $retry$} node[draw, circle,  inner sep=2pt, fill] (s0a) {} node [pos=0.75, above] {\small $1$} (s1);
			\draw (s1) edge[->] node[pos=0.25, above] {\small $send$} node[draw, circle,  inner sep=2pt, fill] (s1a) {} node [pos=0.75, above] {\small $0.1$} (s2);

			\node[coordinate] (l1) [below=1.5 of s1a] {};
			\node[coordinate] (l2) [below=1.5 of s4] {};
			\draw[rounded corners,->] (s1a) -- node[right] {\small $0.9$} (l1) -- (l2) -- (s4);
			
			\draw (s2) edge[->] node[pos=0.25, above] {\small $retry$} node[draw, circle,  inner sep=2pt, fill] (s2a) {} node [pos=0.75, above] {\small $1$} (s3);

			\draw (s3) edge[->] node[pos=0.25, above] {\small $send$} node[draw, circle,  inner sep=2pt, fill] (s3a) {} node [pos=0.75, below] {\small $0.9$} (s4);
			
			\node [nodestyle, node distance=1.1cm, minimum size=0.1cm] (sink) [on grid, above of=s3a] {$\bot$};
			
			\draw (s3a) edge[->] node[pos=0.5, right] {\small $0.1$} (sink);
		\end{tikzpicture}
	}
	\caption{Backwards reachability MDP for $\PTA$ and property $\solprop{\leq}{0.99}{done \land y \leq 10}$.}
	\label{fig:backreach}
\end{figure}
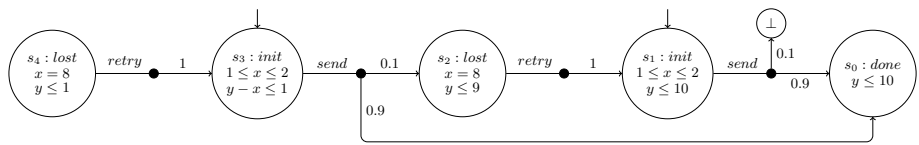

For property $\solprop{\leq}{0.99}{\mathit{done} \land y \leq 10}$, we show the backwards reachability MDP of PTA \PTA in \Cref{fig:backreach}.
The target state is $s_0$.
To obtain its predecessors we apply $\mathit{tpre}$ followed by $\mathit{dpre}$.
We have $s_0 = \mathit{tpre}(s_0)$ (omitting the subscript because it is only needed for until properties).
We then apply $\mathit{dpre}$ for the one incoming edge of $\mathit{done}$, \ie $\mathit{dpre}(\tuple{\mathit{init}, send, \emptyset, \mathit{done}}, \mathit{tpre}(s_0))$.
No clocks are reset, so we only need to intersect zone $y \leq 10$ with guard $x \geq 1$ and invariant $x \leq 2 \land y \leq 24$ of $\mathit{init}$, which yields zone $1 \leq x \leq 2 \land y \leq 10$, making up state $s_1$.
Next, $\mathit{tpre}(s_1) = \tuple{\mathit{init}, x \leq 2 \land y \leq 10 \land y - x \leq 9}$ and for $\mathit{dpre}(\tuple{lost, retry, \set{x}, \mathit{init}}, \mathit{tpre}(s_1))$, we perform a backwards reset on $x$, which will be any clock valuation where $y \leq 9$.
Intersecting with guard and invariant yields $s_2$.
The same principle applies to $s_3$ and $s_4$.
However, since the zone of $s_3$ is a subset of the zone of $s_1$, it inherits the successors of $s_1$.
Therefore, there is an edge directly to $s_0$.
For $s_4$, the predecessor would have an empty zone, which means all predecessors have been discovered and the algorithm terminates.

\section{Implementation}
\label{sec:implementation}
For our implementation of backwards reachability in the \toolset, we followed the pseudocode of \cite{KNSW07} as closely as possible, with some optimizations to eliminate a few obviously redundant calculations.
We use DBMs to represent convex zones, which suffice for maximal reachability probabilities, and lists of DBMs whenever non-convex zones occur for until properties and minimum probabilities or due to disjunctions in constraints.

\begin{figure}[t!]
\centering
\scalebox{0.9}{
\begin{tikzpicture}
	\begin{axis}[
		width=\backopssize,
		height=\backopssize,
	    axis lines=middle,
	    	every axis x label/.style=
				{at={(ticklabel cs: 0.5,0)}, anchor=north},
			every axis y label/.style=
				{at={(ticklabel cs: 0.5,0)}, anchor=east},
			xmin=0,xmax=9,ymin=0,ymax=9,
			xtick distance=1,
			ytick distance=1,
			xlabel=x,
			ylabel=y,
			title={}
	]
		\draw [fill=red, opacity=0.5] (1,2) rectangle (5,5);
		\draw [fill=green, opacity=0.5] (5,3) rectangle (8,8);
		\draw [fill=yellow, opacity=0.5] (3,5) rectangle (5,8);
		\node [text width=0.3cm] at (3, 3.5) {$\zeta_1$};
		\node [text width=0.3cm] at (6.5, 5.5) {$\zeta_2$};
		\node [text width=0.3cm] at (4, 6.5) {$\zeta_3$};
		
		\draw [-stealth,line width=2pt](5.5,0.5) -- (8.5,3.5);
		\node [text width=0.2cm] at (7.5, 1.9) {\tiny dir.};
		\node [text width=0.6cm] at (7.5, 1.3) {\tiny \text{of time}};
	\end{axis}
\end{tikzpicture}
}~
\scalebox{0.9}{
\begin{tikzpicture}
	\begin{axis}[
		width=\backopssize,
		height=\backopssize,
	    axis lines=middle,
	    	every axis x label/.style=
				{at={(ticklabel cs: 0.5,0)}, anchor=north},
			every axis y label/.style=
				{at={(ticklabel cs: 0.5,0)}, anchor=east},
			xmin=0,xmax=9,ymin=0,ymax=9,
			xtick distance=1,
			ytick distance=1,
			xlabel=x,
			title={}
	]
		\draw [fill=red, opacity=0.5] (1,2) rectangle (5,5);
		\draw [fill=green, opacity=0.5] (5,3) rectangle (8,5);
		\draw [fill=yellow, opacity=0.5] (3,5) rectangle (8,8);
		\node [text width=0.3cm] at (3, 3.5) {$\zeta_4$};
		\node [text width=0.3cm] at (6.5, 4) {$\zeta_5$};
		\node [text width=0.3cm] at (5.5, 6.5) {$\zeta_6$};
	\end{axis}
\end{tikzpicture}
}~
\scalebox{0.9}{
\begin{tikzpicture}
	\begin{axis}[
		width=\backopssize,
		height=\backopssize,
	    axis lines=middle,
	    	every axis x label/.style=
				{at={(ticklabel cs: 0.5,0)}, anchor=north},
			every axis y label/.style=
				{at={(ticklabel cs: 0.5,0)}, anchor=east},
			xmin=0,xmax=9,ymin=0,ymax=9,
			xtick distance=1,
			ytick distance=1,
			xlabel=x,
			title={}
	]
		\draw [fill=green, opacity=0.5] (3,5) rectangle (8,8);
		\draw [fill=red, opacity=0.2] (0,0) -- (4,0) -- (4,1) -- (5,1) -- (5,4) -- (7,4) -- (7,6) -- (3,6) --(3,4) -- (1,4) -- (1,1) -- (0,1) -- cycle;
		\draw [fill=yellow, opacity=0.5] (3,5) -- (7,5) -- (6,4) -- (5,4) -- (1,0) -- (0,0) -- (1,1) -- (1,2) --(3,4) -- cycle;
		\node [text width=0.3cm] at (5.5, 6.5) {$\zeta_7$};
		\node [text width=0.3cm] at (4, 1.8) {$\zeta_8$};
		\node [text width=0.3cm] at (4, 4) {$\zeta_9$};
	\end{axis}
\end{tikzpicture}
}
\caption{Non-canonicity of lists of DBMs (left, middle) and $\mathit{tpre}$ complications (right).}
\label{fig:difficulties}
\end{figure}
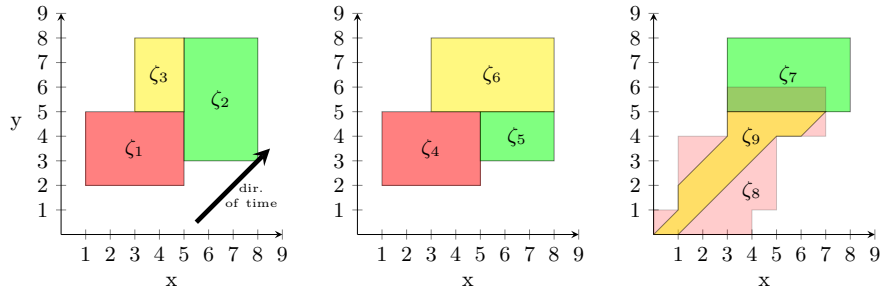

A major difficulty when working with lists of DBMs is the lack of a canonical form as highlighted in the left and middle parts of \Cref{fig:difficulties}, where $[\zeta_1, \zeta_2, \zeta_3]$ contains the same valuations as $[\zeta_4, \zeta_5, \zeta_6]$.
As a consequence, our implementation sometimes makes unnecessary iterations before recognising a fixpoint.
Among the difficulties that we encountered in replicating~\cite{KNSW07} was that the pseudocode for $\mathit{MaxU}$ is missing some operations. 
For example, line 14 adds a new predecessor state to the set of discovered states $Z$, but the journal version~\cite{KNSW07} does not add the corresponding edge to the set of discovered edges, whereas the 2004 paper version~\cite{KNSW04} correctly lists this step (line~15).
Furthermore, \cite{KNSW04} only gives a superficial intuitive explanation of $\mathit{tpre}$, omitting detailed pseudocode and implying that it should be trivial to implement.
As illustrated on the right of \Cref{fig:difficulties} where $tpre_{\tuple{\ell,\zeta_8}}(\tuple{\ell,\zeta_7}) = \tuple{\ell,\zeta_7 \cup \zeta_9}$, however, it is a rather involved procedure.
Finally, the PTA drawn for the CSMA benchmark model in~\cite{KNSW07} appear to be inaccurate:
they include a clock $x$ in some states that is never reset; for the model to make sense, we assumed this to be an error and replaced it with clock~$y$.

\section{Experiments}
\label{sec:Experiments}
In ~\cite{KNSW07}, backwards reachability was benchmarked using models of the CSMA/CD and Firewire protocols for varying values of $c$ and different time bounds. 
Additional benchmarks concerned a property transformation, and compared the state space sizes and generation times to forwards reachability and digital clocks.

We have attempted to replicate these results.
Additionally, we benchmarked the entire model checking procedure, to evaluate the overall competitiveness of backwards reachability.
However, we did not compare against forwards reachability and we did not test the property transformation.
We have tested all the parameter values and properties considered in~\cite{KNSW07} and some additional ones, but we only list an interesting selection below.
Since the original PRISM PTA models are not available, we recreated them from the visual representations in~\cite{KNSW07}. 
To compare with the digital clocks approach, we used the automatic digital clocks transformation in the \textsf{mcsta} tool of the \toolset. 
This might yield models that are slightly different to the manually transformed ones used in the original benchmarks.
All runtimes that we report are in seconds.

\begin{table}[t]
\caption{Graph analysis data for CSMA/CD.}
\label{tab:csma1}
\centering
\scriptsize
\setlength{\tabcolsep}{3pt}
\begin{tabular}{@{}crrrrrrr@{}}
\toprule
 \multirow{3}{*}{\rotatebox[origin=c]{90}{$\mathit{bcmax}~~~\,$}}  &   & \multicolumn{3}{c}{$\solprop{\geq}{1}{done}$}            & \multicolumn{3}{c}{$\solprop{\geq}{\lambda}{done \land z \leq 2000}$ }            \\ \cmidrule(r){3-5} \cmidrule(l){6-8}
      &   &   & \multicolumn{2}{c}{iterations} &   & \multicolumn{2}{c}{iterations} \\ \cmidrule(lr){4-5} \cmidrule(l){7-8} 
 & $c$ & time    & $\mathit{MaxV_{\geq 1}}$          & $\mathit{MaxU_{\geq 1}}$         & time    & $\mathit{MaxV_{\geq 1}}$          & $\mathit{MaxU_{\geq 1}}$         \\ \midrule
  & 50 & 45.2 & 39 & 100 & 30.4 & 39 & 100\\
  & 400 & 18.2 & 7 & 20 & 9.1 & 7 & 20\\
1 & 500 & 20.3 & 6 & 18 & 10.3 & 6 & 18\\
  & 600 & 23.7 & 6 & 18 & 11.4 & 6 & 18\\
  & 700 & 22.6 & 5 & 15 & 10.3 & 5 & 15\\
  & 800 & 24.6 & 5 & 15 & 10.9 & 5 & 15\\
\midrule
  & 50 & 534.4 & 41 & 106 & 394.0 & 41 & 106\\
  & 400 & 103.1 & 7 & 21 & 58.9 & 7 & 21\\
2 & 500 & 110.2 & 6 & 18 & 59.1 & 6 & 18\\
  & 600 & 114.0 & 6 & 18 & 59.1 & 6 & 18\\
  & 700 & 111.2 & 5 & 15 & 55.5 & 5 & 15\\
  & 800 & 111.0 & 5 & 15 & 52.7 & 5 & 15\\
\bottomrule
\end{tabular}
\end{table}

\Cref{tab:csma1} lists the results of the $\mathit{MaxV_{\geq 1}}$ graph analysis necessary to calculate minimal reachability probabilities. 
This replicates~\cite[Table 1]{KNSW07}. 
We tested additional values for $c$, because we noticed that our implementation behaved best for values between $400$ and $800$.
The original sees the best results at $c = 50$, which was relatively slow in our benchmarks.
We suspect that the differences can be mostly attributed to small differences in the recreated model.

\begin{table}[b]
\centering
\caption{States and model checking times of $z.\solprop{\sim}{\lambda}{done \land z {\leq} D}$ for CSMA/CD.}
\label{tab:csma2}
\centering
\setlength{\tabcolsep}{2.5pt}
\scriptsize
\begin{tabular}{@{}crrrrrrrr@{}}
\toprule
 &  & \multicolumn{4}{c}{backwards reachability}        & \multicolumn{3}{c}{digital clocks} \\ \cmidrule(r){3-6} \cmidrule(l){7-9}
 $\mathit{bcmax}$ & $D$ & states (max) & time (max) & states (min) & time (min) & states   & time (max)  & time (min)  \\ \midrule
 & 1200 & 1 & 0.0 & 3 & 13.2 & 10085 & 0.5 & 0.4\\
 & 1600 & 1 & 0.1 & 3 & 15.4 & 10085 & 0.6 & 0.5\\
1 & 2000 & 1025 & 0.1 & 909 & 20.0 & 10085 & 0.7 & 0.6\\
 & 2400 & 1929 & 0.2 & 1741 & 20.9 & 10085 & 0.8 & 0.7\\
 & 2800 & 2833 & 0.5 & 2521 & 22.0 & 10085 & 0.9 & 0.8\\
\midrule
 & 1200 & 1 & 0.8 & 3 & 147.7 & 128553 & 4.7 & 4.5\\
 & 1600 & 1 & 5.9 & 3 & 169.3 & 128553 & 6.1 & 5.9\\
2 & 2000 & 21750 & 9.5 & 28839 & 206.3 & 128553 & 7.7 & 7.4\\
 & 2400 & 27384 & 11.7 & 31783 & 220.2 & 128553 & 9.1 & 8.9\\
 & 2800 & 31744 & 16.2 & 34543 & 239.8 & 128553 & 10.6 & 10.4\\ 
\bottomrule
\end{tabular}
\end{table}

\Cref{tab:csma2} (partly) replicates~\cite[Table~2]{KNSW07}.
We checked $z.\solprop{\sim}{\lambda}{done \land z \leq D}$ and fixed $c$ to $400$.
This is the value of $c$ that produces the best results for our implementation; the value used for the original experiments was not specified.
We notice a significant difference in the number of states. 
Our implementation detects cases where the probability to reach the target is $0$; this happens for lower values of $D$, resulting in very small state counts.
For the other cases, the our implementation can explore more than twice as many states as the original.

One thing to note is that our backwards implementation sometimes outperforms digital clocks.
As opposed to the original work, we benchmarked the time of the entire model checking procedure. 
For maximal probabilities, backwards reachability appears to outperform digital clocks especially for lower time bounds~$D$.
Digital clocks appears to scale better with a higher time bound.
The claim of~\cite{KNSW07} that backwards reachability yields a much smaller state space than digital clocks does not hold any more due to \textsf{mcsta} implementing an unrolling-free technique for time-bounded reachability~\cite{HH16}. 
Such a method is not straightforwardly applicable to backwards reachability. 
We also note that the time benefit of backwards reachability is largely obtained by having a close to optimal value for $c$, which is not trivial to find.

\Cref{tab:firewire1} replicates the data of~\cite[Table 3]{KNSW07}.
These results are closer to the original.
The differences in times can be attributed to us using a much faster 
\begin{wraptable}[11]{r}{7.6cm}
\centering
\vspace{-2.25ex}
\caption{Graph analysis data for the Firewire model.}
\vspace{1.5ex}
\label{tab:firewire1}
\scriptsize
\setlength{\tabcolsep}{3pt}
\begin{tabular}{@{}rrrrrrr@{}}
\toprule
& \multicolumn{3}{c}{$\solprop{\geq}{1}{done}$}            & \multicolumn{3}{c}{$\solprop{\geq}{\lambda}{done \land z \leq 10000}$}            \\ \cmidrule(r){2-4} \cmidrule(l){5-7}
  &  & \multicolumn{2}{c}{iterations} &  & \multicolumn{2}{c}{iterations} \\ \cmidrule(r){3-4} \cmidrule{6-7} 
$c$  & time & $\mathit{MaxV_{\geq 1}}$          & $\mathit{MaxU_{\geq 1}}$         & time & $\mathit{MaxV_{\geq 1}}$          & $\mathit{MaxU_{\geq 1}}$         \\ \midrule
 10& 0.7& 372& 780& 0.2& 372& 780\\
 100& 0.1& 39& 82& 0.0& 39& 82\\
 360& 0.0& 13& 27& 0.0& 13& 27\\
 1670& 0.0& 5& 11& 0.0& 5& 11\\
 2000& 0.0& 4& 9& 0.0& 4& 9\\
 3000& 0.0& 4& 8& 0.0& 4& 8\\
 \multicolumn{7}{c}{$\cdots$}\\
 10000& 0.0& 3& 6& 0.0& 3& 6\\
 \bottomrule
\end{tabular}
\end{wraptable}
processor.
The number of iterations for $\mathit{MaxV_{\geq 1}}$ on property $\solprop{\geq}{1}{done}$ corresponds exactly to the original results.
However, the number of iterations of $MaxU_{\geq 1}$ is different. 
We are unsure of the cause; in particular, our implementation does produce the correct probabilities (compared to digital clocks).
Our only explanation are differences in the model again. 
We also notice that the number of iterations for the time-bounded property $\solprop{\geq}{\lambda}{done \land z \leq 10000}$ in our implementation is equal to the number of iterations for its non-bounded counterpart. 
We came to the conclusion that this is intended as the input to $\mathit{MaxV_{\geq 1}}$ is equal for both properties, with the added time constraint. 
Our implementation keeps the time bound separate from the PTA, so it does not influence the number of iterations.
We speculate that the original prototype or model may have added the time bound to each invariant and guard in order to ensure that it is not violated.

\begin{table}[b]
\centering
\caption{State space sizes and model checking times for the Firewire model.}
\label{tab:firewire2}
\scriptsize
\setlength{\tabcolsep}{3pt}
\begin{tabular}{@{}rrrrrrrr@{}}
\toprule
& \multicolumn{4}{c}{backwards reachability}        & \multicolumn{3}{c}{digital clocks} \\ \cmidrule(r){2-5} \cmidrule(l){6-8}
$D$   & states (max) & time (max) & states (min) & time (min) & states   & time (max)  & time (min)  \\ \midrule
 2000 & 50 & 0.0 & 35 & 0.0 & 7670 & 0.5 & 0.5\\
 4000 & 125 & 0.0 & 83 & 0.0 & 7670 & 0.5 & 1.1\\
 8000 & 455 & 0.0 & 256 & 0.0 & 7670 & 0.5 & 2.1\\
 10000 & 725 & 0.0 & 379 & 0.0 & 7670 & 0.5 & 2.6\\
 20000 & 2655 & 2.0 & 1373 & 0.4 & 7670 & 0.5 & 5.1\\
 40000 & 10300 & 192.9 & 5152 & 21.9 & 7670 & 0.5 & 7.8\\
\bottomrule
\end{tabular}
\end{table}

We also compare to digital clocks using the Firewire model in~\Cref{tab:firewire2} for property $z.\solprop{\sim}{\lambda}{done \land z \leq D}$ as in~\cite[Table 4]{KNSW07}.
We fix $c$ to $4000$.
We observe a similar pattern to the one in \Cref{tab:csma2}, where backwards reachability is faster than digital clocks for lower time bounds. 
Again digital clocks scales better with higher time bounds in both runtime and state-space.
We also observe a larger state space once again of up to three times the number of states in~\cite{KNSW07}.

\section{Conclusion}
\label{sec:Conclusion}
We attempted to replicate the work on backwards reachability for PTA of~\cite{KNSW07} via an implementation in the \toolset that closely follows the original pseudocode.
Our replication was partially successful:
The implementation works and computes the same probabilities as the digital clocks approach where applicable.
However, we see rather different patterns for runtime and iteration counts in several cases, especially when varying $c$ for the CSMA/CD model.

\bibliography{paper}

\begin{thebibliography}{10}
\providecommand{\url}[1]{\texttt{#1}}
\providecommand{\urlprefix}{URL }
\providecommand{\doi}[1]{https://doi.org/#1}

\bibitem{AD94}
Alur, R., Dill, D.L.: A theory of timed automata. Theor. Comput. Sci.
  \textbf{126}(2),  183--235 (1994). \doi{10.1016/0304-3975(94)90010-8}

\bibitem{D89}
Dill, D.L.: Timing assumptions and verification of finite-state concurrent
  systems. In: International Workshop on Automatic Verification Methods for
  Finite State Systems. Lecture Notes in Computer Science, vol.~407, pp.
  197--212. Springer (1989). \doi{10.1007/3-540-52148-8\_17}

\bibitem{HH16}
Hahn, E.M., Hartmanns, A.: A comparison of time- and reward-bounded
  probabilistic model checking techniques. In: Second International Symposium
  on Dependable Software Engineering: Theories, Tools, and Applications,
  ({SETTA}). Lecture Notes in Computer Science, vol.~9984, pp. 85--100 (2016).
  \doi{10.1007/978-3-319-47677-3\_6}

\bibitem{HH14}
Hartmanns, A., Hermanns, H.: The {M}odest {T}oolset: An integrated environment
  for quantitative modelling and verification. In: 20th International
  Conference on Tools and Algorithms for the Construction and Analysis of
  Systems ({TACAS}). vol.~8413, pp. 593--598. Springer (2014).
  \doi{10.1007/978-3-642-54862-8\_51}

\bibitem{HKKS21}
Hartmanns, A., Katoen, J.P., Kohlen, B., Spel, J.: Tweaking the odds in
  probabilistic timed automata. In: 18th International Conference on
  Quantitative Evaluation of Systems ({QEST}). Lecture Notes in Computer
  Science, vol. 12846, pp. 39--58. Springer (2021).
  \doi{10.1007/978-3-030-85172-9\_3}

\bibitem{KNP09}
Kwiatkowska, M.Z., Norman, G., Parker, D.: Stochastic games for verification of
  probabilistic timed automata. In: 7th International Conference on Formal
  Modeling and Analysis of Timed Systems ({FORMATS}). Lecture Notes in Computer
  Science, vol.~5813, pp. 212--227. Springer (2009).
  \doi{10.1007/978-3-642-04368-0\_17}

\bibitem{KNP11}
Kwiatkowska, M.Z., Norman, G., Parker, D.: {PRISM} 4.0: Verification of
  probabilistic real-time systems. In: 23rd International Conference on
  Computer Aided Verification ({CAV}). Lecture Notes in Computer Science,
  vol.~6806, pp. 585--591. Springer (2011). \doi{10.1007/978-3-642-22110-1\_47}

\bibitem{KNPS06}
Kwiatkowska, M.Z., Norman, G., Parker, D., Sproston, J.: Performance analysis
  of probabilistic timed automata using digital clocks. Formal Methods Syst.
  Des.  \textbf{29}(1),  33--78 (2006). \doi{10.1007/s10703-006-0005-2}

\bibitem{KNSS02}
Kwiatkowska, M.Z., Norman, G., Segala, R., Sproston, J.: Automatic verification
  of real-time systems with discrete probability distributions. Theor. Comput.
  Sci.  \textbf{282}(1),  101--150 (2002). \doi{10.1016/S0304-3975(01)00046-9}

\bibitem{KNSW04}
Kwiatkowska, M.Z., Norman, G., Sproston, J., Wang, F.: Symbolic model checking
  for probabilistic timed automata. In: Joint International Conferences on
  Formal Modelling and Analysis of Timed Systems ({FORMATS}) and Formal
  Techniques in Real-Time and Fault-Tolerant Systems ({FTRTFT}). Lecture Notes
  in Computer Science, vol.~3253, pp. 293--308. Springer (2004).
  \doi{10.1007/978-3-540-30206-3\_21}

\bibitem{KNSW07}
Kwiatkowska, M.Z., Norman, G., Sproston, J., Wang, F.: Symbolic model checking
  for probabilistic timed automata. Inf. Comput.  \textbf{205}(7),  1027--1077
  (2007). \doi{10.1016/j.ic.2007.01.004}

\bibitem{P94}
Puterman, M.L.: {M}arkov Decision Processes: Discrete Stochastic Dynamic
  Programming. Wiley Series in Probability and Statistics, Wiley (1994).
  \doi{10.1002/9780470316887}

\end{thebibliography}
\bibliographystyle{splncs04}

\end{document}